\documentclass[aps,prx,onecolumn,superscriptaddress]{revtex4-2}
\usepackage{graphicx}% Include figure files
\usepackage{xcolor}
\usepackage[normalem]{ulem} % For strikethrough

\begin{document}

% Use the \preprint command to place your local institutional report
% number in the upper righthand corner of the title page in preprint mode.
% Multiple \preprint commands are allowed.
% Use the 'preprintnumbers' class option to override journal defaults
% to display numbers if necessary
%\preprint{}

%Title of paper
\title{Large Single-Phonon Optomechanical Coupling between Quantum Dots and Tightly Confined Surface Acoustic Waves in the Quantum Regime}

% repeat the \author .. \affiliation  etc. as needed
% \email, \thanks, \homepage, \altaffiliation all apply to the current
% author. Explanatory text should go in the []'s, actual e-mail
% address or url should go in the {}'s for \email and \homepage.
% Please use the appropriate macro foreach each type of information

% \affiliation command applies to all authors since the last
% \affiliation command. The \affiliation command should follow the
% other information
% \affiliation can be followed by \email, \homepage, \thanks as well.
\author{Ryan A. DeCrescent$^{\ddag}$}
\email{ryan.decrescent@nist.gov}
\affiliation{National Institute of Standards and Technology, Boulder, Colorado 80305, USA}
\thanks{These authors contributed equally to this work.}
\author{Zixuan Wang$^{\ddag}$}
\affiliation{National Institute of Standards and Technology, Boulder, Colorado 80305, USA}
\affiliation{Department of Physics, University of Colorado, Boulder, CO 80309, USA}
\thanks{These authors contributed equally to this work.}
\author{Poolad Imany}
\affiliation{National Institute of Standards and Technology, Boulder, Colorado 80305, USA}
\affiliation{Department of Physics, University of Colorado, Boulder, CO 80309, USA}
\author{Robert C. Boutelle}
\affiliation{National Institute of Standards and Technology, Boulder, Colorado 80305, USA}
\author{Corey A. McDonald}
\affiliation{National Institute of Standards and Technology, Boulder, Colorado 80305, USA}
\affiliation{Department of Physics, University of Colorado, Boulder, CO 80309, USA}
\author{Travis Autry}
\affiliation{National Institute of Standards and Technology, Boulder, Colorado 80305, USA}
\author{John D. Teufel}
\affiliation{National Institute of Standards and Technology, Boulder, Colorado 80305, USA}
\author{Sae Woo Nam}
\affiliation{National Institute of Standards and Technology, Boulder, Colorado 80305, USA}
\author{Richard P. Mirin}
\affiliation{National Institute of Standards and Technology, Boulder, Colorado 80305, USA}
\author{Kevin L. Silverman}
\email{kevin.silverman@nist.gov}
\affiliation{National Institute of Standards and Technology, Boulder, Colorado 80305, USA}

\date{\today}

\begin{abstract}
Surface acoustic waves (SAWs) coupled to quantum dots (QDs), trapped atoms and ions, and point defects have been proposed as quantum transduction platforms, yet the requisite coupling rates and cavity lifetimes have not been experimentally established. Although the interaction mechanism varies, small acoustic cavities with large zero-point motion are required for high efficiencies. We experimentally establish the feasibility of this platform through electro- and opto-mechanical characterization of tightly focusing, single-mode Gaussian SAW cavities at $\sim$3.6 GHz on GaAs. We explore the performance limits of the platform by fabricating SAW cavities with mode volumes approaching 6$\lambda^3$ and linewidths $\leq$1 MHz. Employing strain-coupled single InAs QDs as optomechanical intermediaries, we measure single-phonon optomechanical coupling rates $g_0 \approx 2\pi \times 1.2$ MHz. Sideband scattering rates thus exceed intrinsic phonon loss, indicating the potential for quantum optical readout and transduction of cavity phonon states. To demonstrate the feasibility of this platform for low-noise ground-state quantum transduction, we develop a fiber-based confocal microscope in a dilution refrigerator and perform single-QD resonance fluorescence sideband spectroscopy at mK temperatures. These measurements show conversion between microwave phonons and optical photons with sub-natural linewidths. 
\end{abstract}

% insert suggested keywords - APS authors don't need to do this
%\keywords{}

%\maketitle must follow title, authors, abstract, and keywords
\maketitle

% body of paper here - Use proper section commands
% References should be done using the \cite, \ref, and \label commands
% Put \label in argument of \section for cross-referencing
%\section{\label{}}
\section{Introduction}
Mesoscopic systems that mediate interactions between mechanical motion and light (optomechanical) or electrical signals (electromechanical) have enabled substantial advances in the control, measurement, and transfer of quantum states \cite{oconnell_quantum_2010, chan_laser_2011, teufel_sideband_2011, verhagen_quantum-coherent_2012, palomaki_coherent_2013, safavi-naeini_squeezed_2013, purdy_strong_2013, aspelmeyer_cavity_2014, andrews_bidirectional_2014, balram_coherent_2016, weaver_coherent_2017, satzinger_quantum_2018, higginbotham_harnessing_2018, kotler_direct_2021}. Popular architectures include membranes or phononic crystals – defining the mechanical modes of the system – which capacitively or piezoelectrically couple to electrical circuits and parametrically interact with optical resonators \cite{aspelmeyer_cavity_2014}. Such systems have been used to cool mechanical oscillators to their quantum ground states \cite{teufel_sideband_2011, chan_laser_2011}, produce squeezed states of light \cite{purdy_strong_2013, safavi-naeini_squeezed_2013}, prepare, store, and transfer quantum states \cite{palomaki_coherent_2013}, and to transduce quanta between electrical, mechanical, and optical domains \cite{andrews_bidirectional_2014, shao_microwave--optical_2019, forsch_microwave--optics_2020, mirhosseini_superconducting_2020, han_microwave-optical_2021}. Acoustic modes in bulk structures are also suitable for these purposes \cite{oconnell_quantum_2010, chu_quantum_2017, chu_creation_2018, gokhale_epitaxial_2020} and often offer the benefit of simple fabrication and on-chip integrability, while maintaining long coherence times approaching, and even exceeding, 1 ms. 

Recently, surface acoustic waves (SAWs) --- propagating acoustic waves naturally confined to a medium’s surface --- have emerged as exciting and versatile mechanical modes for quantum systems \cite{schuetz_universal_2015, aref_quantum_2016, weis_interfacing_2018, moores_cavity_2018, delsing_2019_2019}. As electromechanical elements, SAWs efficiently interact piezoelectrically with external microwave circuits --- typically through periodic metallic structures called interdigital transducers (IDTs) --- and strongly couple with superconducting qubits at GHz frequencies \cite{gustafsson_propagating_2014, moores_cavity_2018, satzinger_quantum_2018, dumur_quantum_2021}. When confined within cavities \cite{manenti_surface_2016, shao_phononic_2019}, discreet standing-wave eigenmodes can be selectively and coherently populated \cite{satzinger_quantum_2018}. As optomechanical elements, acoustic waves  parametrically modulate a wide variety of optical systems, e.g., semiconductor quantum dots (QDs) \cite{metcalfe_resolved_2010, weis_optomechanical_2021, imany_quantum_2022}, atoms \cite{schuetz_universal_2015}, defect centers \cite{golter_coupling_2016, weis_interfacing_2018, maity_coherent_2020} and optical cavities \cite{balram_coherent_2016, shao_microwave--optical_2019, forsch_microwave--optics_2020, mirhosseini_superconducting_2020}, and thus effectively mediate electro-optic interactions. SAWs can also be focused, offering opportunities for generating wavelength-scale confinement of phonons in three dimensions \cite{de_lima_focusing_2003, maity_coherent_2020, msall_focusing_2020}. Indeed, because of their innate capacity to couple to a wide variety of optical- and microwave-frequency qubits, SAWs have been recognized as “universal quantum transducers” \cite{schuetz_universal_2015}. 

Within the framework of microwave-to-optical quantum transduction, owing to inherently strong electromechanical interactions, efficient transduction draws attention to optomechanical coupling. A critical threshold to be reached is that where the single-phonon optomechanical coupling rate ($g_0$) exceeds the intrinsic loss rate of the mechanical subsystem. In this case, optomechanical state transfer exceeds the mechanical decoherence rate, and quantum transduction of phonon states becomes feasible. Realized values for $g_0$ vary widely depending on the specific platform,  with best values reported to date lying around 1MHz \cite{aspelmeyer_cavity_2014, han_microwave-optical_2021}. Generally, $g_0$ grows linearly with the zero-point amplitude, $u_{\text{zpm}}$, of the mechanical mode, motivating the development of SAW microcavities with the smallest feasible mode volumes. The ultimate limits of SAW confinement and SAW optomechanical coupling rates with various optical systems have not yet been established.

Here we design, fabricate and characterize high-performance single-mode GaAs SAW cavities at 3.6 GHz with an emphasis on quantum transduction applications. First, we establish several basic design principles for optimized device performance and demonstrate semi-planar cavities with internal quality factors ($Q_i$) exceeding 16,000 and finesse ($F$) exceeding 100. We then explore the confinement limits of SAW phonons by fabricating high-finesse cavities with mode volumes as small as 6$\lambda^3$. We use InAs QDs as local strain probes to internally quantify single-phonon optomechanical couplings of $g_0 \approx 1.2$ MHz in our smallest cavities, exceeding the intrinsic phonon loss ($\lesssim 1$ MHz). The reported performance is comparable to anticipated limits of the platform \cite{schuetz_universal_2015}. Finally, to demonstrate the potential for low-noise ground-state quantum transduction with this platform, we develop a fiber-based confocal microscope in a dilution refrigerator and perform single-QD resonance fluorescence spectroscopy at mK temperatures. Microwave phonons are converted to optical photons that are resonantly and coherently scattered from a single QD. This demonstrates compatibility of this optomechanical system with sample temperatures corresponding to mechanical occupancies of $<<1$ without the need for additional active cooling techniques.

\section{Design Principles of SAW Cavity Systems}
For completeness, we detail the basic design principles of SAW cavities on GaAs. SAW cavities are based on a Fabry-Perot design in which SAWs are confined between two acoustic reflective regions (``mirrors"), resulting in standing-wave strain profiles [Fig. 1a]. We consider SAWs on the (001) GaAs surface propagating predominantly along the [110] direction. SAW mirrors are defined by periodic rectangular etched grooves on the surface with a 50\% duty cycle. This periodicity creates a SAW propagation stop band [Fig. 1b; gray shaded region] \cite{shao_phononic_2019}. The stop-band width and the reflectance of each mirror element increases with the etch depth, resulting in a shorter mirror penetration depth. However, bulk-scattering losses also increase with etch depth \cite{schuetz_universal_2015}. We find that a 20 nm etch depth (2.5\% of the SAW wavelength, $\lambda$) [Fig. 1b; vertical dashed line] results in a good balance between confinement and bulk loss. A typical penetration depth, $L_p$, into the mirror region for this etch depth is $L_p \approx 7$ $\mu$m for SAWs of wavelength 800 nm at 3.6 GHz. Stop-band widths are approximately 100 MHz.

%%%%%%%% FIGURE 1 %%%%%%%
\begin{figure*}
      \includegraphics[width=0.75\textwidth]{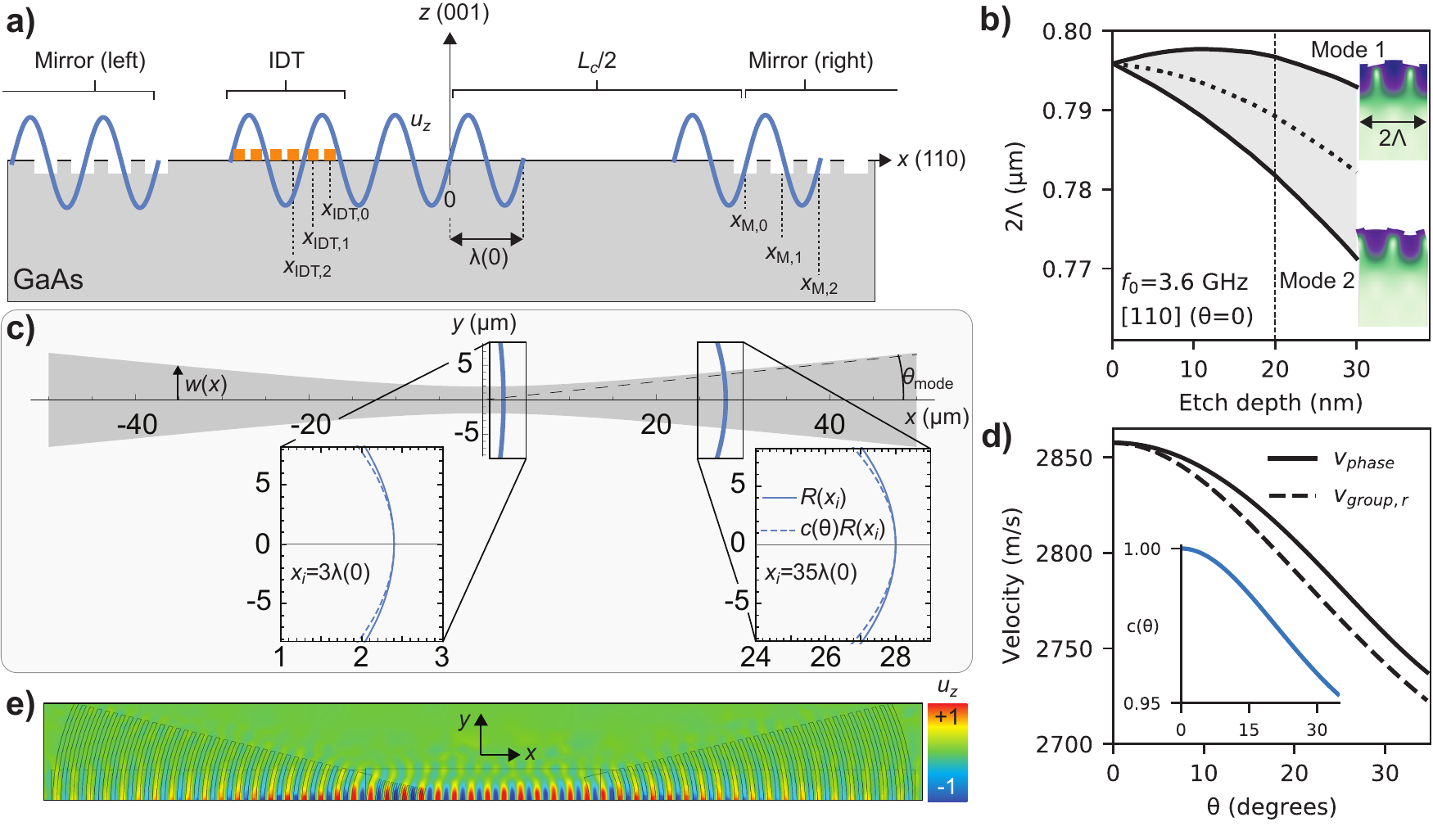}
	\caption{	
	Design principles of Gaussian focusing SAW cavities. 
	(\textbf{a}) Cross-sectional illustration of the SAW cavity with the standing-wave pattern along the $x$ [110] direction and the substrate normal in the $z$ [001] direction. White regions represent etched mirror elements. IDT fingers are represented by gold rectangles. Blue sinusoids illustrate the out-of-plane displacement, $u_z$, of the SAW standing wave. The $x$ positions of the mirror (`M') and IDT elements are represented below each element.
	(\textbf{b}) Surface-wave eigenmodes of the periodic etched-groove system defining the SAW mirrors. The spacing between the etched grooves is $\Lambda$. The mode’s spatial periodicity, $2\Lambda$, is calculated at a fixed frequency, $f_0$ (here, $f_0$=3.6 GHz). Mode 1 and Mode 2 (solid black curves) become nondegenerate as the mirror etch is introduced. For periodicities $2\Lambda$ between the two modes (filled gray region), SAWs cannot propagate and are reflected. Fabricated devices use $2\Lambda$ values at the center of the gap (dotted black curve). 
	(\textbf{c}) Illustration of the Gaussian beam profile in the $x$-$y$ plane. Gray region designates the $1/e$ field region, i.e., beam waist (Gaussian half-widths) $w(x)$, for $\theta_{\text{mode}}=7.5^\circ$. Two phase fronts, at $x_i=3\lambda(0)$ [left] and $x_i=35\lambda(0)$ [right], are indicated by blue curves. Insets: phase-front curvatures with (dashed) and without (solid) curvature corrections. 
	(\textbf{d}) SAW phase velocity (solid black curve) and the radial component of the group velocity (dashed black line) calculated as a function of the geometric angle, $\theta$, with respect to the [110] direction on the bare GaAs (001) surface. Inset: $\theta$-dependent radial correction function [$c(\theta)=v_{\text{group},r}(\theta)/v_{\text{group},r}(0)$] for Gaussian mode curvature on the GaAs (001) surface. 
	(\textbf{e}) Finite-element calculation of the out-of-plane displacement field, $u_z$, in a focusing cavity, illustrating the Gaussian mode profile described above.
	}
\label{fig:1}
\end{figure*}

Cavity design begins by specifying a desired operating frequency, $f_0$. For this frequency, the mirror periodicity, $\Lambda$, is chosen to lie in the center of the stop band [Fig. 1b; dotted curve]. The SAW wavelength for this frequency along the [110] direction of the bare GaAs surface, $\lambda(0)$, is calculated from the phase velocity $v_{\text{phase}}(\theta)$ at propagation angle $\theta=0$. In this work, the angle $\theta$ is referenced with respect to the GaAs [110] direction. The cavity length, $L_c$, is chosen to be an integer multiple of $\lambda(0)$, $L_c=n\lambda(0)$, with the mirror etches lying at nodes of the $z$-displacement of the standing wave. SAWs are excited using an interdigital transducer (IDT) fabricated within the cavity. We use a ``double finger” IDT design in order to eliminate reflections from the individual IDT fingers \cite{aref_quantum_2016, moores_cavity_2018}. The IDT is strategically positioned so that the periodic electrical potential applied to the IDT overlaps maximally with the periodic potential of the SAW standing wave [Fig. 1a]. 

For ``planar" (no curvature) cavities, these basic 1D design principles are simply extended an arbitrary distance along the $y$ axis. For focusing cavities, the structures follow the phase fronts of a 2D Gaussian beam in the $x-y$ plane (Fig. 1c). The mode structure is uniquely defined by a ``mode angle", $\theta_{\text{mode}}$ [dashed black line], which describes the asymptotic behavior of the beam half-width, $w(x)$ [filled gray region], far away from the beam focus at $x$=0. Ignoring wave velocity anisotropy, the phase fronts of this beam intersecting the $x$ axis as position $x_i$ are described by circular arcs of radius $R(x_i)=x_i[1+(x_R/x_i)^2]$, where $x_R=\lambda(0)/(\pi\theta_{\text{mode}}^2)$ is the Rayleigh length of the beam [solid blue curves]. SAW velocity anisotropy on the GaAs (001) surface modifies the mode curvature [dashed blue curves] according to $v_{\text{group},r}(\theta)/v_{\text{group},r}(0)$ where $v_{\text{group},r}(\theta)$ is the radial component of the group velocity at angle $\theta$ [Fig 1d] \cite{maznev_anisotropic_2003, msall_focusing_2020}. The mirror and IDT structures essentially must follow the phase fronts of this mode profile. Numerical calculations of the SAW cavity strain fields [Fig. 1e] illustrate the described mode shape. Our fabricated devices take these details fully into account without any further approximations. Additionally, we take into account variations in the SAW wavelength under the IDT and consequent corrections to the total cavity length. Further calculation and design details are provided in Supplementary Information Section A. Device fabrication is described in ref. \cite{imany_quantum_2022}.

\section{Characterization of Cavity Loss Mechanisms}
In the limit of no focusing, SAWs propagate with a single momentum component and losses due to the pseudo-confinement of off-[110] propagating SAWs are negligible \cite{kuok_angular_2001}. In this case we can quantify losses due to propagation and bulk scattering from mirrors. Figure 2a shows a scanning electron microscope (SEM) image of a fabricated planar SAW cavity device. The 10-period IDT (i.e., spanning 10 SAW wavelengths in the $x$ dimension) is positioned at the center of a cavity of length $L_c=50\lambda(0) \approx 41.29$ $\mu$m. The width ($y$ dimension) of the IDT is $W=33$ $\mu$m. The small number of IDT periods offers a broad bandwidth ($\sim$500 MHz) that enables us to excite all cavity modes within the mirrors’ stop band ($\sim$100 MHz) and allows the fabrication of short cavities. Even with few IDT periods, we have achieved electromechanical conversion efficiencies up to 90\% in planar cavities.

%%%%%%%% FIGURE 2 %%%%%%%
\begin{figure*}
      \includegraphics[width=0.75\textwidth]{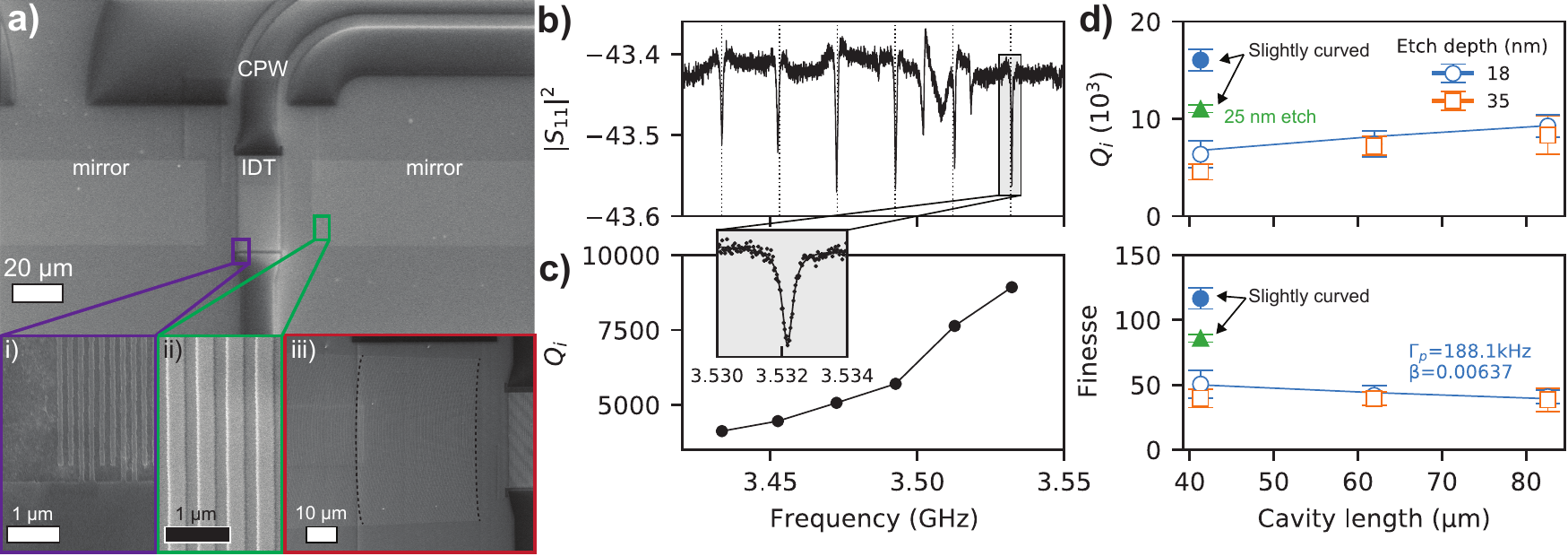}
	\caption{	
	Microwave characterization of planar and semi-planar devices. 
	(\textbf{a})  Upper panel: SEM image of a device with 10 IDT periods inside a planar cavity of length $L_c=50\lambda(0)\approx41.29$ $\mu$m. Mirror and IDT regions are identified with white text. CPW: coplanar waveguide. Lower panels: i-ii) Details of planar cavity regions specified by colored squares. iii) SEM image of a similar cavity intentionally fabricated with slightly curved mirrors. Curvature is accentuated by a dashed black curve. 
	(\textbf{b}) Microwave reflectance [$10\log(|S_{11}|^2)$] of a planar SAW cavity with length $L_c=75\lambda(0)\approx61.93$ $\mu$m. Low-frequency background oscillations in data have been removed for clarity in post-processing; note that this filtering process slightly affects the shape of narrow spectral features. Vertical dashed lines indicate uniform mode spacing with an FSR of 19.67 MHz. Inset shows the Lorentzian model fit (solid curve) of the data (markers) around the 6th cavity mode (indicated by gray rectangle in panel b). 
	(\textbf{c}) Internal quality factor, $Q_i$, as a function of mode frequency in the same device. 
	(\textbf{d}) Internal quality factor, $Q_i$ (upper panel), and finesse, $F$ (lower panel), for a series of devices with different cavity lengths (horizontal axis) and etch depths (see legend). Open markers (error bars) represent the mean (standard deviation) of all modes identified in the $|S_{11}|$ spectrum of the respective device. Results from two different slightly curved cavities are included (solid markers); the green triangle is from a device with a 25-nm etch depth. Solid blue curve: Fit of 18-nm etch planar cavity data to the model in Eqns. 1-2 with $\Gamma_p=2\pi\times$ 188.1 kHz and $\beta$=0.00637. All measurements were performed at 20 mK.
	}
\label{fig:2}
\end{figure*}

Microwave reflection measurements ($S_{11}$) at 20 mK reveal a series of evenly spaced cavity modes [Fig. 2b; vertical dotted lines] where the reflectance $|S_{11}|^2$ is sharply reduced. Two additional sharp features in the spectrum likely arise from other transverse cavity modes or external microwave resonances in our setup; they do not affect our conclusions. The reflection around each cavity mode is well fit by a single Lorentzian lineshape [Fig. 2b; inset]. Our fitting procedure allows us to simultaneously extract the internal quality factor of each mode, $Q_i$, and the electromechanical coupling rate, $\kappa_{\text{EM}}$ \cite{aref_quantum_2016}. In this work we focus on $Q_i$ as this informs on the internal mechanical performance of the cavities. For this specific device ($L_c=50\lambda(0)\approx41.29$ $\mu$m; mirror etch depth 18 nm), we measure $Q_i$s of approximately 10,000 at the highest frequencies [Fig. 2c], corresponding to linewidths (full width at half maximum; FWHM) of $\approx350$ kHz. Interestingly, $Q_i$ generally increases as the cavity mode approaches the high-frequency edge of the mirror’s stop band. This effect is consistent with numerical calculations and likely originates from decreased bulk scattering as the standing-wave field profile in the mirror regions approaches high-frequency mode structure shown in Fig. 1b. This effect can be exploited to optimize cavity lifetimes in single-mode cavities by positioning the mode frequency near the top of the stop band. 
 
Comparing similar results across various cavity design parameters (e.g., cavity lengths and etch depths) allows us to identify the dominant loss mechanisms. We consider propagation loss ($\Gamma_p$), mirror scattering ($\Gamma_m$), and diffraction losses. Each source is modeled in terms of known geometrical parameters and the SAW velocity \cite{aref_quantum_2016, schuetz_universal_2015}. For the moment ignoring diffraction, expressions for $Q_i$ and $F$ for a cavity mode of frequency $f_0$ are given by [Supplementary Information Section B]

\begin{equation}
Q_i = \left( \frac{\Gamma_p}{f_0} + 2\beta \frac{v_{\text{phase}}(0) / [2(L_C+2L_p)] }{f_0} \right)^{-1}
\end{equation}
\begin{equation}
F = \left( \frac{\Gamma_p}{v_{\text{phase}}(0) / [2(L_C+2L_p)] } + 2\beta \right)^{-1}
\end{equation}
\begin{equation}
\Gamma_m = 2\beta \frac{v_{\text{phase}}(0)}{2(L_C+2L_p)}
\end{equation}

\noindent where $\beta$ ($0<\beta<1$) is a scalar fit parameter that describes the proportional phonon loss per reflection from the mirrors. Fig. 2d shows the measured $Q_i$ (upper panel) and $F$ (lower panel) as a function of cavity length for several mirror etch depths. Fits to Eqns. 1-2 reveal that propagation and mirror losses are approximately balanced for the cavity lengths investigated here [Fig 2d; solid curves]; specifically, absolute loss rates from propagation and mirror losses are $\Gamma_p=2\pi\times$ 188.1 kHz and $\Gamma_m=2\pi\times$ 196.1 kHz (for the 83 $\mu$m cavity, 18 nm etch). Increased etch depths lead to stronger phonon confinement but we find that both $Q_i$ and $F$ are generally reduced as etch depths exceed 18 nm. 

Diffraction loss is difficult to quantify by similar methods, but analytical estimates \cite{aref_quantum_2016, manenti_surface_2016} suggest that for planar devices, diffraction loss rates ($\sim$$2\pi\times 270$ kHz) are comparable to measured propagation loss rates even for these wide cavities with $W\approx37\lambda(0)$. Including a constant diffraction-loss term in Eqns. 1-2 should affect the inferred propagation and mirror loss rates, but the overall conclusion is expected to remain the same: propagation and mirror losses are roughly balanced for these cavity lengths. Further reductions to cavity lengths are needed to enter a regime where performance is dominated by mirror losses, simultaneously increasing finesse while reducing mode volumes.

A slight inward mirror curvature can effectively mitigate diffraction losses, substantially increasing $Q$s in Fabry-Perot cavities. Fig. 2a (panel iii) shows a fabricated SAW cavity with mirrors adopting a very slight curvature. The radius of curvature is approximately 10 times the cavity half-length, and the overall cavity dimensions are comparable to that in Fig. 1a. Nonetheless, for an 18 nm etch depth, $Q_i$ ($F$) increases by more than a factor of two, to a value of 16,500 (120). To our knowledge, these are the highest reported values for $Q_i$ and $F$ in SAW cavities on GaAs reported at these frequencies to date. 

\section{Focusing SAW Cavities for Improved Optomechanics}
The analyses above indicate that single-phonon interactions can be significantly improved by reducing cavity lengths and strategically accounting for diffraction. Previous work has demonstrated focusing of SAWs on GaAs \cite{msall_focusing_2020}, AlAs/Diamond \cite{maity_coherent_2020}, and AlN \cite{whiteley_spinphonon_2019} substrates. We first assess the limits of SAW focusing by fabricating a series of devices with conservative cavity lengths and a range of mode angles, which produce tightly focused beams [Fig. 3a]. The mode angles for this set of devices are $7.5^\circ$, $15.0^\circ$ and $22.5^\circ$, corresponding to designed focal-plane mode waists, $w_0$, of $2.4\lambda(0)$ [1.9 $\mu$m], $1.2\lambda(0)$ [0.95 $\mu$m], and $0.8\lambda(0)$ [0.64 $\mu$m], respectively. The total cavity length is $L_c=82\lambda(0)$ [$\sim$63 $\mu$m] for all devices shown here. Note that the mirror and IDT structures exceed the designed mode width to ensure that scattering from structure edges is reduced; nonetheless, optical mapping of the strain field verifies that the mode profile is as designed \cite{imany_quantum_2022}. Microwave reflectance measurements [Fig. 3b] reveal three primary cavity modes within the mirror stop bands. The measured $Q_i$ for the central mode [Fig. 3c] is clearly reduced as SAWs are more tightly focused. Since the device designs within this series are otherwise identical, we attribute this reduction in $Q_i$ largely to bulk losses inherent to pseudo-SAWs propagating at oblique angles with respect to the GaAs [110] axis \cite{kuok_angular_2001}.

Experimentally, we see that $Q_i$ decreases by only a factor of $\sim$3 between large planar cavities and short focusing cavities [Figs. 2d and 3c]. Nonetheless, optomechanical performance is expected to improve significantly. The optomechanical coupling rate is proportional to the zero-point displacement field at the focus, $u_{\text{zpm}}(0,0)$. Therefore, for a Gaussian focusing cavity, $g_0 \sim w_0^{-1} (w_0L_c)^{-1/2} \sim \theta_{\text{mode}}^{3/2} / L_c^{1/2}$, which is easily increased by orders of magnitude for short, tight-focusing cavities compared to long planar cavities (e.g., Fig. 2). System cooperativities and efficiencies are expected to scale roughly as $g_0^2Q_i$, motivating the development of short, tightly focusing SAW cavities. 

%%%%%%%% FIGURE 3 %%%%%%%
\begin{figure*}
      \includegraphics[width=0.75\textwidth]{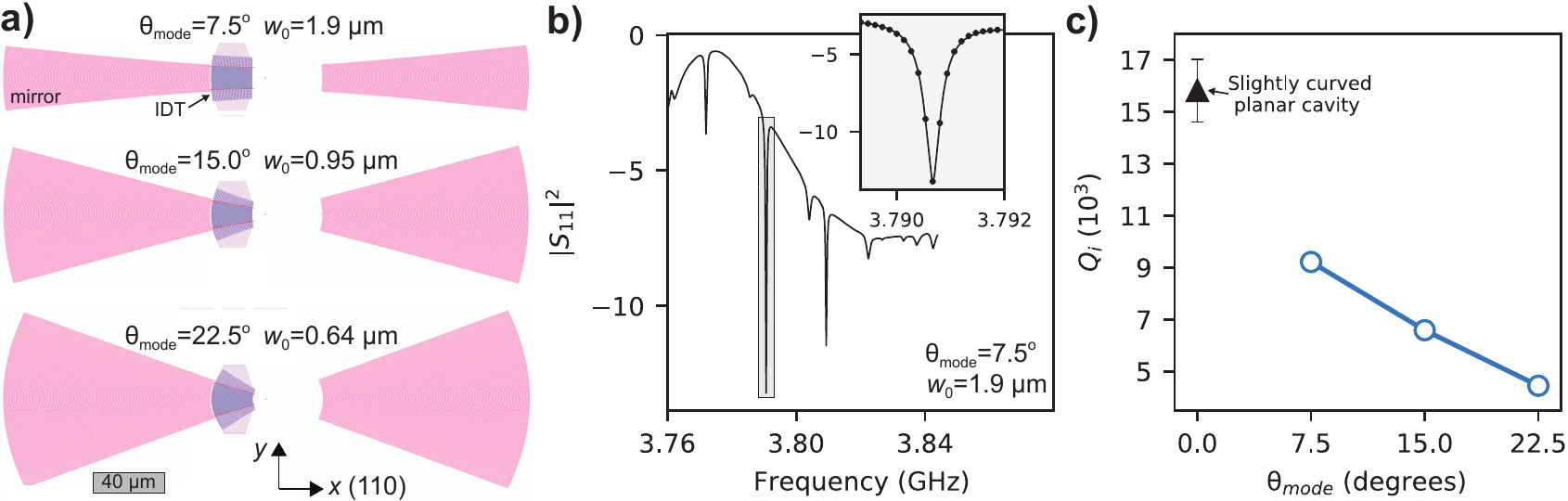}
	\caption{	
	Microwave characterization of Gaussian focusing SAW cavities. 
	(\textbf{a}) Device structure of three devices with different mode angles, $\theta_{\text{mode}}$. Designed focal-plane beam waists (half-widths), $w_0$, correspond approximately to $2.4\lambda(0)$ [1.9 $\mu$m], $1.2\lambda(0)$ [0.95 $\mu$m], and $0.8\lambda(0)$ [0.64 $\mu$m]. All devices here have total cavity length $L_c=82\lambda(0)\approx63$ $\mu$m, and 30 IDT periods. Pink: SAW mirrors. Dark purple: IDTs. Light purple: microwave waveguide traces (truncated here for illustration purposes). Gray scale bar is 40 $\mu$m wide. 
	(\textbf{b}) Microwave reflectance [$10\log(|S_{11}|^2)$] of the $\theta_{\text{mode}}=7.5^\circ$ cavity shown in panel a. Inset: Close-up of the central cavity mode around 3.791 GHz (designated by gray rectangle). Markers: data. Solid curve: Lorentzian fit. The extracted linewidth (FWHM in power, $|S_{11}|^2$) is 411 kHz. c) Measured internal quality factors, $Q_i$, as a function of mode angle, $\theta_{\text{mode}},$ for the series of devices shown in panel a (blue circles). To accentuate the trend, $Q_i$ of the slightly curved planar cavity with 18 nm mirror etch depth shown in Fig. 2d is included (black triangle) at $\theta_{\text{mode}}$=0. All measurements were performed at 20 mK.
	}
\label{fig:3}
\end{figure*}

\section{SAW Microcavities and Applications to Quantum Transduction}
Upon reducing cavity lengths to $\sim28\lambda(0)$ [$\sim$22 $\mu$m], devices show only a single mode within the mirror stop bands, indicating an FSR approaching or exceeding $\sim$40 MHz. Linewidths (in power, $|S_{11}|^2$) remain smaller than 1 MHz even for tightly focusing cavities with designed beam waists of $w_0\leq1.0$ $\mu$m. Fig. 4a illustrates the geometry of a fabricated small-mode-volume SAW cavity with $L_c\approx19\lambda(0)$ [18.37 $\mu$m]. A single cavity mode with a FWHM of 739 kHz at $f_0$=3.550 GHz ($Q_i=4800$) is observed as a sharp dip in the $|S_{11}|^2$ spectrum (Fig. 4b). Calculated estimates of the cavity FSR (44.2 MHz) in combination with this measured linewidth indicate $F>60$. 

The short IDTs with few periods offer relatively weak electromechanical coupling at the current stage. Nonetheless, these small cavities were fabricated on substrates containing InAs QDs, and we exploit the inherently strong optomechanical interaction between SAWs and QDs to optically quantify the mechanical performance, optomechanical coupling rates, and single-phonon displacements. These measurements directly relate to properties fundamental to applications in quantum microwave-to-optical transduction.

We measure photoluminescence (PL) spectra from single QDs subject to the internal SAW cavity strain field at a sample temperature of 5K. QDs are excited with a non-resonant optical pump, and the PL spectrum from a single QD is acquired by scanning a voltage-tunable optical filter around the QD's emission frequency [Supplementary Information Section C]. The QD exciton energy is modulated by the SAW's strain field, and the luminescence spectrum acquires a series of phonon-mediated sidebands \cite{metcalfe_resolved_2010, weis_optomechanical_2021, imany_quantum_2022}. QDs thus act as local strain probes where the relative sideband intensities correspond to the local strain field. The modulation index, $\chi$, uniquely describes this relative sideband intensity, and directly relates to the single-phonon coupling rate via $\chi=2g_0\sqrt{\bar{n}}/\omega_m$, where $\bar{n}$ is the steady-state cavity phonon number [Supplementary Information Section D] \cite{golter_coupling_2016, imany_quantum_2022}. Fig. 4c shows example spectra from a single QD (emitting at $\sim$925 nm, FWHM$\approx$1.6 GHz) with the SAW cavity driven on resonance (black) and 3 MHz away from resonance (red) at a constant incident microwave power of $-50.6$ dBm ($\sim$8.7 nW). On resonance, spectra correspond to a modulation index $\chi$=2.29. Previous work required significantly greater than $-5$ dBm (0.3 mW) \cite{wigger_resonance_2021} to +14 dBm (25 mW) \cite{weis_optomechanical_2021} to achieve comparable modulation strengths. That is, the strong phonon confinement here provides greater than $10^4-10^6$ enhancement in the local strain intensity (per incident microwave photon) around the QD compared to SAWs launched on bare GaAs. This is particularly remarkable considering the very few ($\sim$10) periods of short ($\sim$5 $\mu$m) IDT fingers used here compared to typically $\sim$100 periods of $\sim$25-40 $\mu$m long IDT fingers. Identical measurements at microwave frequencies spanning the SAW cavity resonance [Fig. 4d] reveal the very strong frequency dependence of the local strain field. These curves correspond to a local SAW intensity ($\sim$$\chi^2$) contrast of $>$100 when driving on resonance vs. off resonance and reveal the high-quality SAW resonance. In general, this illustrates a remarkably sensitive method for measuring internal mechanical performance of the phonon cavity. These measurements also allow us to quantify optomechanical coupling rates and transduction efficiencies.

%%%%%%%% FIGURE 4 %%%%%%%
\begin{figure*}
      \includegraphics[width=0.75\textwidth]{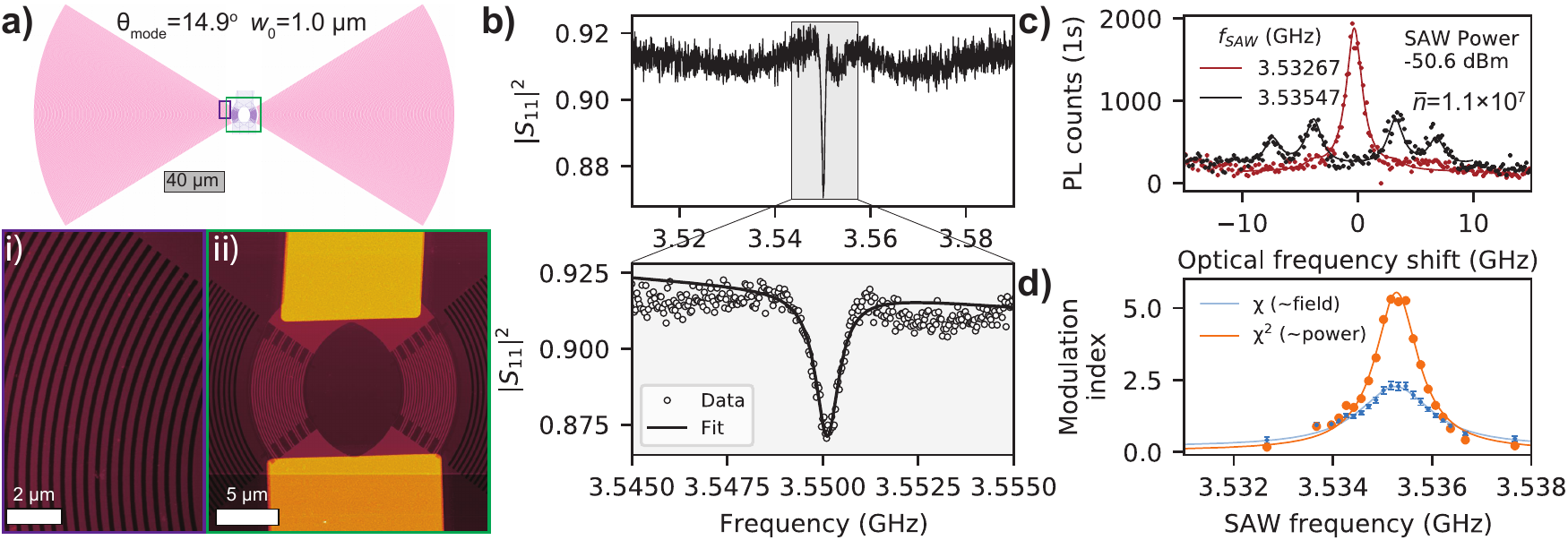}
	\caption{	
	Quantifying high optomechanical coupling rates between SAWs and QDs in small-mode-volume cavities.
	(\textbf{a}) Device structure of a cavity with $\theta_{\text{mode}}=14.9^\circ$, $w_0=1.0$ $\mu$m [$\sim$1.25$\lambda(0)$], and $L_c=18.4$ $\mu$m. This device has 10 total IDT periods, divided into two symmetric portions across the cavity’s center. Pink: SAW mirrors. Dark purple: IDTs. Light purple: microwave waveguide traces (truncated here for illustration purposes). Gray scale bar is 40 $\mu$m wide. Inset panels (i) and (ii) are false-color atomic force microscope images of a fabricated device in regions specified by purple and green rectangles, respectively. 
	(\textbf{b}) Microwave reflectance [$10\log(|S_{11}|^2)$] of the cavity shown in panel a at 1.7K. Upper panel: Spectrum showing only a single cavity mode over an 80 MHz range. Lower panel: Spectrum showing data (open circles) and Lorentzian fit (solid curve) of the single mode around 3.550 GHz. The extracted linewidth (FWHM in power, $|S_{11}|^2$) is 739 kHz. 
	(\textbf{c}) PL spectra of a single QD at the cavity center when driving the SAW cavity on resonance ($f_{\text{SAW}}=3.53547$ GHz; black) and off resonance (3.53267 GHz; red) at a constant microwave power of $-50.6$ dBm. Markers: data. Solid curves: fits. Frequency is referenced with respect to the QD’s center frequency. The cavity used for this measurement has $L_c=22.5$ $\mu$m, $\theta_{\text{mode}}=16.6^\circ$, and 20 total IDT periods (distinct from the device used or panels a-b). The estimated on-resonance steady-state phonon number in the cavity is specified by $\bar{n}$ [Supplementary Information Section D]. 
	(\textbf{d}) Modulation index, $\chi$ (blue), and modulation index squared, $\chi^2$ (orange), as a function of SAW driving frequency at $-50.6$ dBm. $\chi$ ($\chi^2$) is proportional to the local strain field (intensity) at the QD’s location. Error bars on blue markers correspond to $3\sigma$ confidence intervals derived from the fitting algorithm. Solid curves are Lorentzian fits to the data showing a 1 MHz cavity FWHM. Optical measurements shown in panels c-d were performed at 5K.
	}
\label{fig:4}
\end{figure*}

The efficient transduction of microwave-frequency electromagnetic signals to the optical domain is a critical step toward the realization of large-scale quantum networks \cite{kimble_quantum_2008, han_microwave-optical_2021}. Self-assembled InAs/GaAs QDs mediate this microwave-to-optical transduction process wherein a series of phonon-mediated sidebands are imprinted on a resonant and coherent optical pump \cite{metcalfe_resolved_2010, wigger_resonance-fluorescence_2021, imany_quantum_2022, horiuchi_quantum_2022}. Indeed, our system is analogous to an optomechanical cavity wherein the QD plays the role of the optical cavity. The coupling rate and efficiency have been shown to be enhanced by orders of magnitude by positioning the QD within a SAW cavity, with single-phonon optomechanical coupling rates, $g_0$, reaching $2\pi \times 42$ kHz \cite{imany_quantum_2022, horiuchi_quantum_2022} \cite{aspelmeyer_cavity_2014, han_microwave-optical_2021}.
 
The SAW microcavity [Fig. 4a] exhibits exceptionally large and improved single-phonon coupling rates. From our modulation data shown in Fig. 4c-d, we derive $g_0=2\pi \times 1.2$ MHz for this specific cavity [Supplementary Information Section D]. A reasonable error range for this estimate is $\pm0.25$ MHz, arising largely from uncertainty in the equilibrium cavity phonon number [Supplementary Information Section D]. Critically, $g_0$ exceeds the intrinsic phonon loss. The current improvement largely originates from the significantly decreased mode volume, tight focusing, and favorable QD position in the standing-wave SAW field. From this experimental estimate of $g_0$ and literature values for the deformation potential ($G\approx6.5\times10^{14}$ Hz) \cite{metcalfe_resolved_2010}, we estimate a zero-point phonon displacement of $u_{\text{zpm}}\approx1.3$ fm [Supplementary Information Section D]\cite{schuetz_universal_2015}. Purely theoretical estimates of $u_{\text{zpm}}\approx1$ fm and $g_0\approx0.82$ MHz are obtained for this specific cavity when approximately taking into account the focusing SAW mode structure \cite{schuetz_universal_2015, imany_quantum_2022} [Supplementary Information Section D]. It is interesting to note that the smallest feasible cavity (area $\sim 1 \lambda^2$) is expected to have $u_{\text{zpm}} \approx 2.5$ fm and $g_0 \approx 2\pi\times 2$ MHz \cite{schuetz_universal_2015, metcalfe_resolved_2010}; our experimentally derived values for $g_0$ and $u_{\text{zpm}}$ are approaching these anticipated limits. 

An ideal quantum transduction platform must be able to transfer a quantum state between microwave and optical domains efficiently, without additional thermal noise \cite{forsch_microwave--optics_2020, han_microwave-optical_2021}, and with low background photon count rates. At 3.6 GHz, our mechanical resonators are expected to contain fewer than one thermal phonon for temperatures below 250 mK. To this end, we develop a fiber-based confocal microscope system in a dilution refrigerator and perform resonance fluorescence sideband spectroscopy on a single QD in a SAW cavity [Fig. 5; Supplementary Information Section E]. The device is held on a sample stage mounted to the dilution refrigerator's mixing chamber held at 125 mK [Fig. 5a]. The single QD is illuminated by a continuous-wave, diffraction-limited optical pump, and a SAW cavity mode at frequency $f_0$=3.658 GHz is coherently driven with an external microwave source. To demonstrate the feasibility of this platform for microwave-to-optical quantum transduction, we tune the optical pump frequency by $+f_0$ (blue detuned) from the QD’s bare exciton frequency and collect and count inelastically scattered photons through a tunable Fabry-Perot etalon with a $\sim$25 MHz linewidth. Reflected pump light is rejected by $\sim$$10^3$ via polarization [Supplementary Information Section E] and an additional $\sim$$10^5$ via one-pass spectral filtering [Fig. 5b]. Single-phonon events are detected as photons scattered predominantly at the QD’s resonance energy [Fig. 5c]; very weak scattering at $+2f_0$ indicates that single phonons are preferentially added to the cavity in each scattering event [Supplementary Information Section E] and verifies operation in the resolved-sideband limit \cite{metcalfe_resolved_2010, imany_quantum_2022}. Scattering linewidths ($\sim$25 MHz) are significantly narrower than the lifetime-limited QD width ($\sim$200 MHz), indicating coherent photon scattering from the single-photon emitter \cite{matthiesen_subnatural_2012}. Photon count rates $\gtrsim$2 kHz are measured; accounting for system transmission efficiencies ($\sim$0.01) and light-trapping effects in the device structure ($\sim$0.06 collection efficiency), we estimate a photon collection rate of $\sim$200 kHz and a total sideband scattering rate of $\gtrsim$3.3 MHz for this steady-state phonon occupation under microwave driving [Supplementary Information Section E]. Indeed, phonons can be \emph{removed} from the cavity at comparable rates under red-detuned pumping [Supplementary Information Section E]. 

The device used for this specific measurement is similar to the planar cavity shown in Fig. 2, thus requiring relatively high microwave driving powers compared to the tightly focusing cavities shown in Figs. 3-4. The base temperature of 125 mK during the measurements shown in Fig. 5 is higher than typical base temperatures ($\sim$20 mK) due to additional optical and electrical dissipation in the system and sub-optimal thermal contact between the sample stage and the mixing chamber of the dilution refrigerator.

We extrapolate our results to single-phonon levels and estimate an optomechanical transduction efficiency of $\eta_{om}\sim10^{-10}$ for the SAW microcavity illustrated in Fig. 4c. This agrees well with the anticipated $10^5$ improvement with respect to previous results \cite{imany_quantum_2022} due to the improvement in $g_0$ alone (i.e., 1.2 MHz vs. 3 kHz in ref. \cite{imany_quantum_2022}). A discrepancy on the order of 10 likely arises from differences in the cavity lifetimes. We anticipate an additional several orders of magnitude improvement with moderate design modifications to both optical and mechanical subsystems. Open photonic bullseye structures \cite{davanco_circular_2011, decrescent_semicircular_2022} or lensed optical cavities \cite{tomm_bright_2021} can improve photon collection efficiencies by over $100\times$ with respect to the current devices. We can apply these same design principles to fabricate Lamb-wave resonators which exhibit significantly longer phonon lifetimes ($\sim$10$\times$) and improved phonon spatial confinement ($\sim$2-5$\times$) \cite{chou_measurements_2020}. Exploiting resonance effects between the exciton fine structure splitting and the mechanical mode may provide an additional $\sim 50\times$ improvement \cite{kepesidis_phonon_2013}. Impedance-matching networks may significantly improve electromechanical efficiencies \cite{aref_quantum_2016, wu_microwave--optical_2020}.

\section{Discussion and Conclusions}
Distributing entanglement between computing nodes, e.g. as per the DLCZ protocol \cite{sangouard_quantum_2011}, is expected to be the most viable initial use for this system. In this case, quantum state transfer must occur before mechanical decoherence and without added thermal noise, but the condition that successful transfer occurs once every cavity lifetime is relaxed. For our mechanical resonators at 3.6 GHz, a typical dilution refrigerator temperature of 20 mK corresponds to a mechanical occupancy of $10^{-4}$. We wish to emphasize that single QDs efficiently scatter photons at low optical pump powers of $\sim10$ nW, compared to $\sim\mu$W or $\sim$mW pump powers typically used in optical cavities \cite{jiang_efficient_2020}. Further, in contrast to photonic crystal cavities \cite{balram_coherent_2016, forsch_microwave--optics_2020, jiang_efficient_2020}, the optical field in our system is not highly localized in the host material nor near etched interfaces thus making our system less susceptible to optical heating, highlighting the potential for effective low-noise transduction without active laser cooling techniques. The QDs' spectral properties can be improved using charge-control structures \cite{tomm_bright_2021}. Background levels in our resonance fluorescence sideband measurements can be improved by more strategically preparing the polarization state of the incident beam \cite{phoenix_full_2020} and by using additional spectral filters in the collection line \cite{maccabe_nano-acoustic_2020}. Lower base temperatures at our dilution refrigerator’s sample stage are readily achievable through better thermal contact to the heavy sample stage and by eliminating all unnecessary sources of heat dissipation during the measurement. These results suggest that quantum optomechanical measurements with SAW cavities and QDs is a plausible near-future endeavor.

%%%%%%%% FIGURE 5 %%%%%%%
\begin{figure}
      \includegraphics[width=3in]{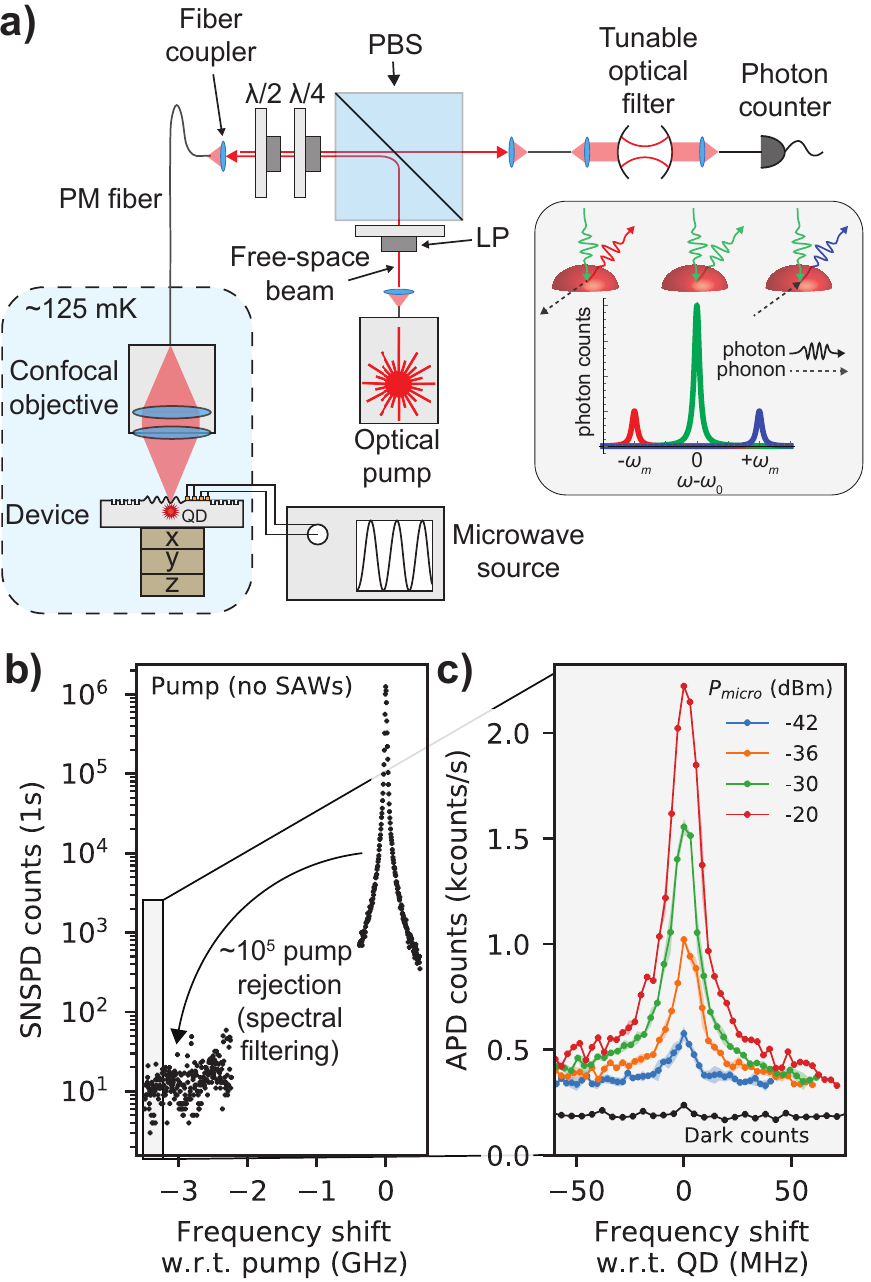}
	\caption{	
	Microwave-to-optical transduction from a QD in a SAW cavity. 
	(\textbf{a}) Schematic of a fiber-based confocal microscope for single-QD resonance fluorescence spectroscopy in a dilution refrigerator. A custom-built objective is fixed above a position-controlled SAW cavity device. The objective couples the tip of a single-mode polarization-maintaining (PM) optical fiber to a diffraction-limited focused spot at the device surface. A wavelength-tunable laser transmits through a series of polarizing optics ($\lambda$/4 and $\lambda$/2 waveplates) before coupling into a polarization-maintaining fiber. Reflected pump light and resonance fluorescence is collected through the same fiber. Collected photons transmit through a polarizing beam splitter (PBS) for pump rejection [Supplementary Information Section E] and through a voltage-tunable Fabry-Perot etalon (25 MHz linewidth) before being counted by either a superconducting nanowire single photon detector (SNSPD) or an avalanche photodiode (APD). Scattering spectra are acquired by scanning the etalon over the sideband frequencies. The SAW cavity is driven on resonance by a coherent external microwave source. Inset: Schematic QD energy spectrum when coupled to SAWs under weak microwave driving showing a no-phonon scattering process (green) and single-phonon scattering processes (red/blue). LP: linear polarizer. 
	(\textbf{b}) Collected pump light is spectrally rejected by $\sim$5 orders of magnitude with the etalon tuned approximately 3.5 GHz from the pump. Combined with $\sim10^3$ polarization rejection, this results in $\sim10^8$ total pump rejection at first-order sideband frequencies. 
	(\textbf{c}) Resonance fluorescence sideband spectra as a function of frequency detuning for various microwave driving powers ($P_{\text{micro}}$). The pump frequency is tuned by $+f_0$ from the QD’s bare exciton frequency (blue sideband pumping); photons are collected around the QD’s bare exciton frequency (set to 0 frequency). The dark count level of the APD is specified by black markers. Filled colored regions represent $1\sigma$ variation of count levels over several measurements. The device used for this specific measurement is similar to the planar cavity shown in Fig. 2. All measurements were performed at 125 mK.
	}
\label{fig:5}
\end{figure}

% If you have acknowledgments, this puts in the proper section head.
%\begin{acknowledgments}
%\end{acknowledgments}

\newpage
%%%%%%% SUPP INFO%%%%%%%%%%%%%%%%
\section*{Supplementary Information: Large Single-Phonon Optomechanical Coupling between Quantum Dots and Tightly Confined Surface Acoustic Waves in the Quantum Regime}
\subsection{SAW velocity calculations and further details of SAW cavity design}
SAW phase and group velocities as a function of propagation direction on the GaAs [001] surface were calculated using a commercially available finite element method solver. We calculate wave velocities independently for three distinct regions: the bare GaAs surface, etched mirror regions, and under the superconducting interdigital transducers. For all calculations, the two-dimensional (2D) environment consisted of a rectangle of height ($y$ dimension) $5\lambda$ and a width ($x$ dimension) of $1\lambda$ ($\lambda$=820 nm). A lower rectangle perfectly matched layer (PML) region was included to account for power dissipation into the substrate bulk. GaAs piezomechanical matrix elements were taken from ref. \cite{balram_coherent_2016}. These piezomechanical matrices correspond to a coordinate system $\{x,y,z\}$ aligned with the GaAs (100), (010), and (001) directions. Periodic boundary conditions were applied to both material boundaries in the $x$ dimension. Solid mechanics and electrostatics modules were used with piezoelectricity multiphysics coupling. A rotated coordinate system was applied to the rectangle in order to align GaAs [001] with the $y$ axis and GaAs [110] with the $x$ axis. Eigenfrequency calculations were performed around 3.5 GHz. SAW modes were identified by their inherently large quality factors and by visualizing the resulting mode structures. Such calculations were performed at each of a series of angles by rotating the coordinate system by an angle $\theta$ about the $y$ axis (the [001] direction) with respect to the [110] direction. The phase velocity, $v_{\text{phase}}(\theta)$, was calculated from each result via $v_{\text{phase}}(\theta)=\lambda f_0(\theta)$ where $f_0(\theta)$ is the eigenfrequency calculated at angle $\theta$. 

The phase velocity on bare GaAs was calculated as described immediately above without further modifications; the results are shown in Fig. 1d of the main text and in Fig. 6. To calculate the group velocity, we first define the frequency-momentum dispersion relationship for SAWs on GaAs [001] according to $\omega(\mathbf{k})=\omega(k,\phi)=v_{\text{phase}}(\phi)k$, where $\mathbf{k}=k\{\cos\phi, \sin\phi\}$. The group velocity vector, $\mathbf{v}_{\text{group}}$, is defined as usual by $\mathbf{v}_{\text{group}}=\nabla_\mathbf{k}\omega(\mathbf{k})$. To take the derivatives, we model the phase velocity over the entire angular range as $v_{\text{phase}}(\phi)=v_0+v_a \sin[(2\pi/p)\phi]^2$, where $\phi$ is the $k$-space angle coordinate defined with respect to the [110] direction. The rotational symmetry properties of GaAs [001] require $p=\pi$. However, best agreement with numerical calculations requires $v_0=2866$ m/s, $v_a=-149$ m/s, and $p=1.06\pi$. The model fit to the numerical data is shown in Fig. 6. To define the anisotropy-based curvature correction at geometric angle $\theta$ (inset of Fig. 1b), we take the radial component of the group velocity at angle $\phi(\theta)$: $v_{\text{group},r}(\theta)=v_{\text{group},x}[\phi(\theta)]\cos\theta+v_{\text{group},y}[\phi(\theta)]\sin\theta$. Here, $\phi(\theta)$ is an angle in $k$-space that differs from the angle $\theta$ due to the fact that the group velocity vector does not point radially inward/outward on an anisotropic surface \cite{de_lima_focusing_2003}. The function $\phi(\theta)$ is the solution to the equation derived by setting $\mathbf{v}_{\text{group}}(\phi)$ to be parallel to $\mathbf{v}_{\text{phase}}(\theta) = v_{\text{phase}}(\theta)\hat{r}$. Specifically, $\phi(\theta)$ is the solution to: $\tan^{-1}[v_{\text{group},y}(\phi) / v_{\text{group},x}(\phi)]=\theta$. At this angle, $v_{\text{group},r}(\theta)=|\mathbf{v}_{\text{group}}(\theta)|$.

For the mirror etch regions, two rectangular regions of width $\lambda/4$, depth $d$, and spacing $\Lambda=\lambda/2$ were removed from the simulation environment (symmetrically about the center of the simulation environment). Calculations were performed at each of a series of values of $d$. Two modes were identified from each result, corresponding to the two modes illustrated in Fig. 1b of the main text, and the phase velocity of each mode was calculated using $v_{\text{phase}}(\theta)=\lambda f_0(\theta)$. For IDT regions, four rectangles of width $\lambda/8$, height 20 nm, and spacing $\lambda/4$ were added to the simulation environment. Niobium material parameters were applied to the four additional rectangles. 

The phase velocity differs in each region. How the phase velocity is applied to the bare GaAs regions and in mirror regions is described in the main text. Specifically, defining the cavity center to lie at $x$=0, the $x$-axis position of the first mirror element (index $m$=0), $x_{M,0}$, is simply $x_{M,0}=L_c/2$. Subsequent mirror elements are positioned incrementally along the $x$ axis: $x_{M,m}=x_{M,0}+m\Lambda$ [Fig. 1a]. An equivalent mirror structure is defined on the opposite side of the cavity with mirror elements at $x_{M,m}=-x_{M,0}-m\Lambda$. For the standing-wave mode with $n$ (integer) wavelengths, the IDT fingers are positioned at $x_{\text{IDT},m}=(-1/8+m/4)\lambda(0)$ for arbitrary integers $m$ which keep $x_{\text{IDT},m}$ within the cavity. These structures are illustrated in Fig. 1a. In our fabricated devices, we also take into account changes in the phase velocity under the IDT region. Specifically, for a given design frequency, the SAW wavelength under the IDTs is reduced by an amount $\delta\lambda$, and the total cavity length is thus reduced by an amount $N_{\text{IDT}}\delta\lambda$ for an IDT with $N_{\text{IDT}}$ periods. That is, the total number of standing-wave wavelengths across the entire cavity length, $L_c$, is preserved in this procedure. 

\begin{figure}
      \includegraphics[width=2.5in]{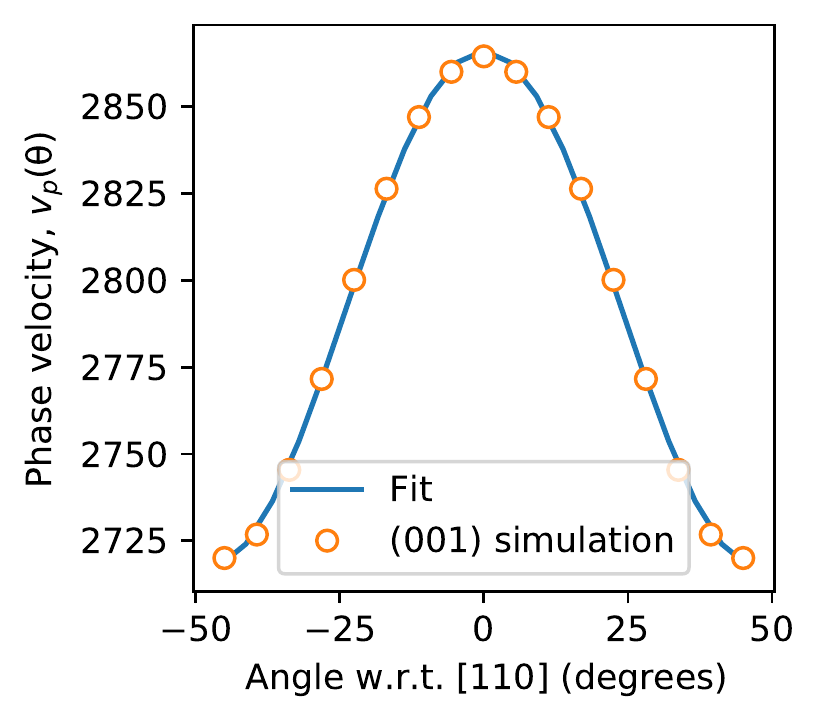}
	\caption{	
	Orange markers: Calculated SAW phase velocity as a function of angle with respect to the [110] direction on a GaAs (001) surface. Solid blue curve: Fit to the calculated values using the expression specified in Supplementary Information Section A.
		}
\end{figure}

\subsection{Expressions for the internal quality factor and finesse for SAW cavities}
We consider two loss mechanisms and derive expressions for the quality factor and finesse for a cavity of length $L_c$. The loss rate due to propagation is independent of the geometry of the SAW cavities, and is thus essentially a constant. Define the constant rate of propagation loss as $\Gamma_p$. The loss rate due to mirror scattering, $\Gamma_m$, depends on the number of mirror reflections per unit time, schematically: $\Gamma_m$=(loss/time)=(loss/reflection)$\times$(reflections/time). We define a constant factor $\beta$ to describe the phonon loss per reflection. The number of reflections per time depends on the SAW velocity and the total cavity length $L_c+2L_p$ via: (reflections/time)=$1/[(L_c+2L_p)/v_{\text{phase}}(0)]$. The factor $v_{\text{phase}}(0)/(L_c+2L_p)$ is identified as 2$\times$FSR where FSR is the cavity’s free spectral range. The loss rates add at a specific cavity frequency $f_0$, and so the total internal quality factor, $Q_i$, is $Q_i$=$f_0 / (\Gamma_p+\Gamma_m)$. The finesse is defined by replacing $f_0$ by the FSR in the previous expression. The model shown in Fig. 2b was fit to the measured $F$. We take a constant value $L_p$=7 $\mu$m in our model, which was derived from independent measurements of the cavity FSR for an 18 nm etch depth.

\subsection{QD photoluminescence sideband measurements}
Single-QD photoluminescence (PL) measurements were performed using the system described in ref. \cite{imany_quantum_2022}. The SAW cavity device was held at $\sim$5 K in an evacuated cryostation. QDs near the SAW cavity’s center were optically pumped with a diffraction-limited focused beam of a 632 nm laser. Collected pump light was rejected using a long-pass optical filter. Photons from a single QD were isolated using a 2 nm bandpass filter and a voltage-tunable Fabry-Perot etalon filter with a $\sim$600 MHz bandwidth. The SAW cavity was actively driven at a specified microwave power and frequency using an external vector network analyzer (VNA). PL spectra were collected at each microwave frequency while keeping the microwave power constant. The spectrum corresponding to each microwave frequency was independently fit for the modulation index, $\chi$, using the expression available in ref. \cite{metcalfe_resolved_2010}. The QD width and amplitude and the SAW frequency were constrained to independently measured values. A parameter corresponding to a constant background count level was included in the fit function. Example fits for all recorded spectra are shown in Fig. 7.

\begin{figure}
      \includegraphics[width=3in]{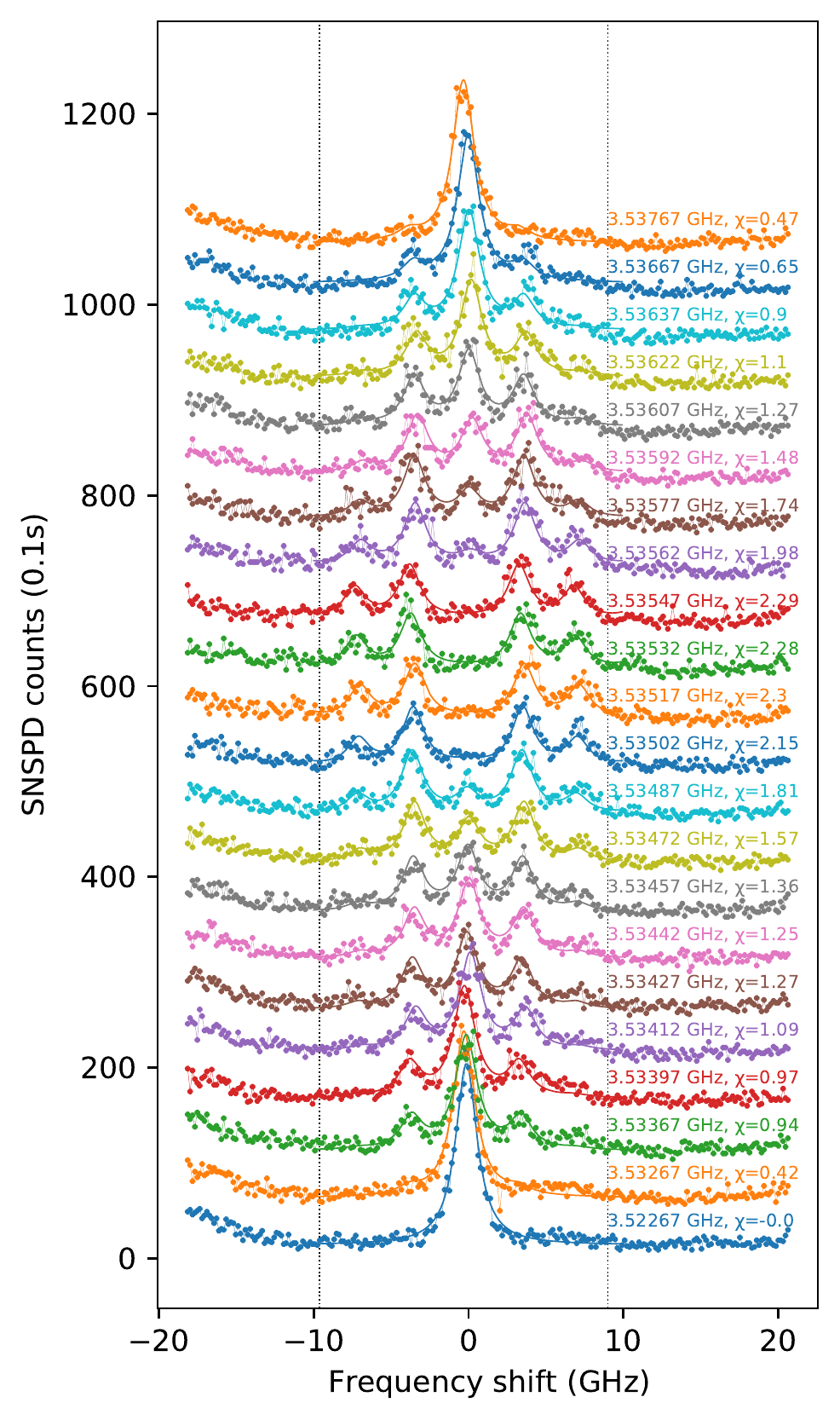}
	\caption{	
	Single-QD PL spectra measured under a constant SAW cavity microwave driving power ($-50.6$ dBm calibrated) and various microwave driving frequencies (specified by text above each spectrum). Each spectrum has been translated vertically for clarity. Solid curves are fits to each spectrum using the expression from ref. \cite{metcalfe_resolved_2010}. Modulation index for each fit is also specified in text above each spectrum. Frequency is referenced with respect to the QD’s center frequency. The horizontal fit range is designated by vertical dotted black lines.
		}
\end{figure}

\subsection{Estimating $g_0$, $u_{\text{zpm}}$, and transduction efficiencies}
Our optomechanical system consisting of a QD in a strain field is analogous to more common optomechanical cavities wherein the QD essentially plays the role of the optical cavity. The QD's instantaneous energy is sensitive to the local strain field, providing the definition for $g_0$:
\begin{equation}
g_0 = [2\pi/\lambda(0)] G u_{\text{zpm}}
\end{equation}

\noindent Here, $G$ is the strain susceptibility (deformation potential) of the QD exciton: $G=d\omega_{\text{QD}} / d\epsilon|_{\epsilon=0}$ where $\epsilon$ is the strain field. The exact value of $G$ in this scalar treatment depends on the specific position of the QD within the SAW strain field; we use 6.5$\times$$10^{14}$ Hz \cite{metcalfe_resolved_2010}. The effective Rabi rate for single-phonon scattering, $\Omega'$, is given by 
\begin{equation}
\Omega'=g_0 \Omega_0/\omega_m
\end{equation}

\noindent where $\Omega_0$ is the bare generalized Rabi frequency between the QD and the detuned pump electric field \cite{golter_coupling_2016}. The factor $g_0/\omega_m$ is equivalent to the Lamb-Dicke parameter in the theory of optical scattering from trapped ions. A practical upper limit to $\Omega_0$ is estimated by ensuring that the steady-state QD population remains low under detuned pumping. We estimate $\Omega_0 \approx 5.05$ GHz under 3.6 GHz detuned pumping for a QD with a linewidth of 500 MHz. 

The modulation index is defined as 
\begin{equation}
\chi = 2g_0\sqrt{\bar{n}}/\omega_m
\end{equation}

\noindent where $\bar{n}$ is the steady-state phonon number \cite{golter_coupling_2016, imany_quantum_2022}. We can thus derive single-phonon optomechanical coupling rates ($g_0$) from the fit modulation index ($\chi$). The square-root dependence of $\chi$ on $\bar{n}$ corresponds to a linear dependence on the SAW field. We estimate $\bar{n}$ from from the incident microwave power, the measured cavity linewidth, and the estimated microwave coupling using 
\begin{equation}
\bar{n} = \frac{\eta\times P_{\text{micro}}}{\hbar \omega_m} \frac{1}{2\pi \Gamma_{\text{mech}}/2}
\end{equation}

\noindent where $\hbar$ is the reduced Planck’s constant, $P_{\text{micro}}$ is the microwave power incident on the IDTs, $\omega_m = 2\pi f_0$ is the SAW frequency in angular units, $\eta$ is the proportion of incident microwave power coupled into the cavity (determined from the microwave reflectance spectrum, $|S_{11}|^2$), and $\Gamma_{\text{mech}}/2\pi$ is the internal loss rate of the SAW cavity mode. To calculate $g_0$, we use the following values, derived from the \emph{in situ} microwave reflectance spectrum at 5K (Fig. 8; top panel): $f_0$=3.536 GHz; $\Gamma_{\text{mech}}=2\pi\times 1$ MHz; $\eta$=0.0091 (corresponding to a 0.04 dB variation in the $|S_{11}|^2$ spectrum); $P_{\text{micro}}$ (calibrated)=$-50.6$ dBm. The estimate $\Gamma_{\text{mech}}=2\pi\times1$ MHz is verified precisely by the internal strain measurement shown in Figs. 4c-d. The result is $g_0=1.2$ MHz $\pm$ 0.25 MHz. Uncertainty arises largely from the estimates of the equilibrium phonon number due to $\sim\pm$1.5 dB uncertainty in attenuation from coaxial cables, connections, wirebonds, and co-planar waveguides in our setup and devices. A dispersive lineshape, likely arising from increased electrical resistance in the optical cryostat setup at 5K, complicates a precise estimate of the external electromechanical coupling rate, but we assume a reasonable uncertainty of $\sim\pm$0.01 dB. 

\begin{figure}
      \includegraphics[width=3in]{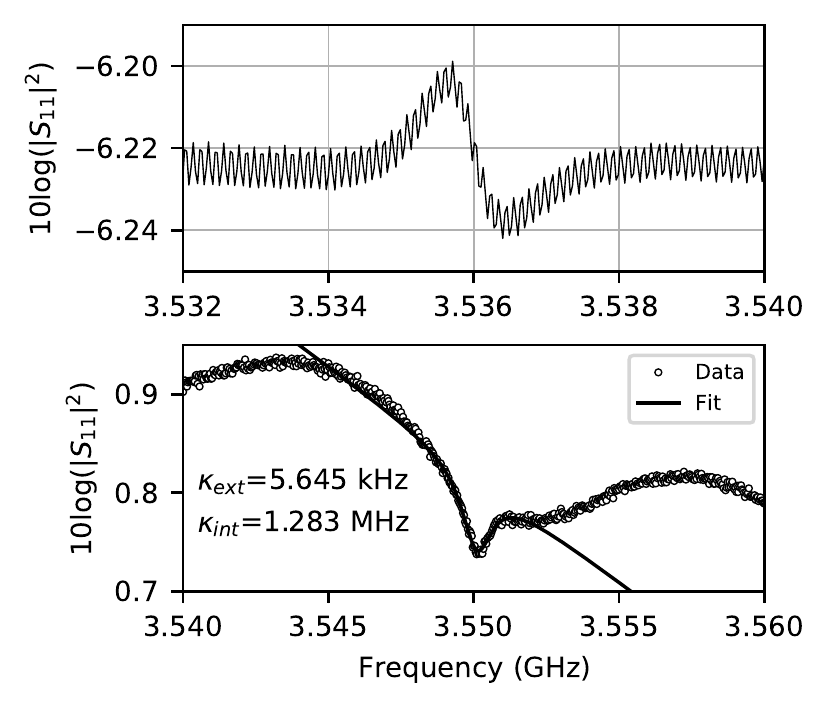}
	\caption{	
	Top: In situ microwave reflectance spectrum [$10\log(|S_{11}|^2)$] from the device used for the optical measurements shown in Figs. 4c-d at $\sim$5K. Slow background variations have been subtracted. Bottom: Microwave reflectance spectrum from a similar device in a different cryostat at 1.6K. The fit internal loss rate ($\kappa_{\text{int}}$) and external coupling rate ($\kappa_{\text{ext}}$) are specified in the plot.
		}
\end{figure}

We alternatively estimate $g_0$ using the microwave characterization of a very similar device in a different cryostat at 1.6K (Fig. 8; bottom panel) via
\begin{equation}
\bar{n} = \frac{4 P_{\text{micro}}}{\hbar \omega_m} \frac{\kappa_{\text{ext}}}{\kappa^2}
\end{equation}

\noindent where $\kappa_{\text{ext}}$ is the external electromechanical coupling rate and $\kappa = \kappa_{\text{int}} + \kappa_{\text{ext}}$ where $\kappa_{\text{int}}$ is the internal loss rate of the cavity \cite{jiang_efficient_2020}. We derive $\kappa_{\text{ext}} = 2\pi\times 5.645$ kHz from a Lorentzian fit to the microwave reflectance data [Fig. 8; lower panel]. An internal loss rate $\kappa_{\text{int}}=2\pi\times1$ MHz is assumed from the internal strain characterization shown in Figs 4c-d. The result is $g_0 =1.158$ MHz.  

The zero-point displacement amplitude, $u_{\text{zpm}}$, was estimated from our experimentally derived $g_0$ using Eqn. 4. Purely theoretical estimates of the zero-point displacement were also derived via $u_{\text{zpm}}=2$ fm$/\sqrt{A [\mu\text{m}^2]}$ \cite{schuetz_universal_2015} where $A [\mu\text{m}^2]$ is the mode area expressed in units of $\mu$m. We calculated $A$ from $A=V/\lambda$ where $V$ is the SAW cavity mode volume \cite{schuetz_universal_2015}. Mode volume was calculated using methods detailed in ref. \cite{imany_quantum_2022}. For a cavity length of 18 $\mu$m, a penetration length of 7 $\mu$m and a mode angle of 0.26 radians, the results are $V\approx3\mu$m$^3 \approx 6 \lambda^3$ and $u_{\text{zpm}} \approx 1$ fm and $g_0 \approx 0.81$ MHz. Differences between theoretical and experimental estimates may arise from the position dependence of the QD-SAW coupling in the non-uniform SAW strain distribution. 

Optomechanical transduction efficiencies, $\eta_{om}$ were derived for the SAW cavity illustrated in Fig. 4c,d assuming -50.6 dBm of incident microwave power to achieve maximal first-order sideband scattering when driving the cavity on resonance, the measured electromechanical coupling efficiency (0.91\%), and a typical sideband photon count rate of 2000 counts per second.

\subsection{Resonance Fluorescence Sideband Spectroscopy in a Dilution Refrigerator}
Resonance fluorescence sideband spectroscopy was performed using a custom-built fiber-based confocal microscope setup with polarization and spectral filtering, illustrated in Fig. 5. The basic optical setup is motivated by the designs described in refs. \cite{kuhlmann_dark-field_2013, imany_quantum_2022}. The critical modification is the replacement of the sample objective with a single-mode polarization-maintaining (PM) optical fiber. Additionally, we use both a $\lambda/4$ plate and a $\lambda/2$ plate to prepare the polarization state of the excitation laser beam before coupling into the PM fiber. We use 850 nm elliptical-clad PM optical fiber with a 100-$\mu$m coating and no external jacket. This fiber is fed into our dilution refrigerator (DR) through a fixed vacuum feedthrough and connected to a custom-built objective with 0.5 numerical aperture comprising two aspheric lenses. The fiber tip is positioned in the focal plane of one of the lenses. The objective is fixed above the sample/device. The sample/device is mounted to the sample post of a position-controllable cryo positioning stage and positioned in the focal plane of the second lens of the objective. Basically, the fiber tip is coupled to a diffraction-limited spot at the sample surface. The IDT in the device is contacted electrically via coplanar waveguides to coaxial cables to an external microwave source. 

To perform sideband spectroscopy, reflected and collected pump light must be maximally rejected. For polarization rejection, the basic goal is to prepare the incident beam in one of the polarization-maintaining modes of the PM fiber. While there are typically two orthogonal linearly polarized modes that satisfy this criteria, we find that the significant temperature variation along the fiber between the mixing chamber ($\sim$100 mK) and ambient ($\sim$293 K) causes polarization rotation and ellipticity. Though unpredictable, we can compensate for it \emph{in situ} using the external wave plates while keeping the fiber tip fixed. The frequency of a single QD is first identified, and then the pump frequency is chosen depending on the desired scattering process. With this pump frequency fixed, we monitor the reflection signal and rotate the external polarizing optics until the signal is maximally rejected. We typically find relatively stable behavior with $\sim$$10^3-10^4$ polarization rejection, with occasional spontaneous transient hops up to $\sim$$10^2$ rejection, presumably due to thermal fluctuations or vibrations along the fiber [Fig. 9]. 

This collected signal containing both residual reflected pump light (polarization-reduced) and light scattered by the QD is transmitted through a voltage-tunable Fabry-Perot etalon with a $\sim$25 MHz linewidth (full width at half max) before being fiber-coupled once more and sent to a photon counter. The desired phonon scattering process is selected by tuning the etalon transmission frequency with respect to the pump. We estimate a total system transmission efficiency of 0.01 between the collection optics and the photon counter. We use transfer-matrix methods to account for light-trapping effects in our device structure \cite{schuller_orientation_2013, iyer_unidirectional_2020} and calculate collection efficiencies of $\lesssim$0.06 into our NA=0.5 objective from QDs emitting in a $\lambda$-thick cavity between two DBRs. Therefore, from measured count rates on the APD, we estimate total sideband scattering rates exceeding $\sim$3 MHz under these specific experimental conditions.

For the measurement shown in Fig. 5 and Fig. 10, the optical pump power incident on the sample interface was $\lesssim$ 300 nW at 911.8949 nm (1st-order blue sideband of the QD). This pump power is larger than typical $\sim$10 nW required pump powers due to sub-optimal alignment of the sample/QD with the pump/collection optics. An electromechanical coupling efficiency of $\sim$60\% was measured for this particular cavity mode. A cavity phonon population of 16.6$\times$$10^9$ (2.6$\times$$10^{12}$) is estimated when driving the cavity at $-42$ dBm ($-20$ dBm). Photons were collected around the QD’s center energy. Under blue-detuned pumping at $+f_0$ (Fig. 10; blue curves) the rate of scattering to $+2f_0$ with respect to the QD is relatively weak, illustrating the resonant enhancement of scattering to the QD’s center frequency and verifying that our system lies in the resolved sideband limit. A similar spectrum is acquired under red-detuned pumping at $-f_0$ (Fig. 10; red curve), highlighting the potential for sideband cooling in this system. Observable differences in count rates between red-and blue-detuned pumping arise from pump spectral misalignment with the red sideband and from contributions from other quasi-resonant QDs in the pump spot.

\begin{figure}
      \includegraphics[width=3.5in]{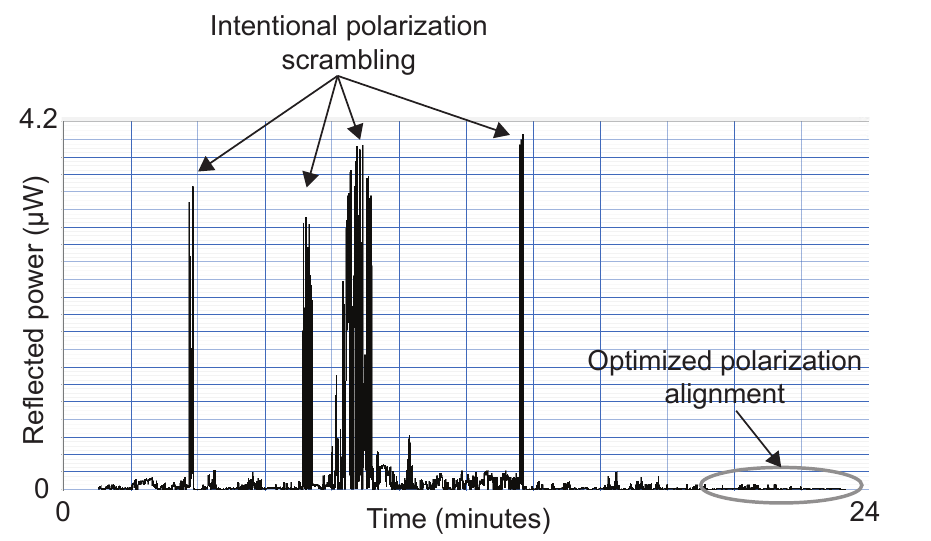}
	\caption{	
	Reflected and collected pump light over a $\sim$24 minute time interval during polarization alignment of the two wave plates. The four power spikes indicated by arrows were caused by intentional polarization scrambling. The peak power of $\sim$4 $\mu$W is approximately half of the total collected reflected pump power because of the linear polarizers in the collection path. The final $\sim$3-4 minutes in this plot (designated by a gray ellipse) were recorded with the wave plates at their optimized angles. The typical power in this region is $\sim$10 nW, corresponding to $\sim$$10^3$ polarization rejection and showing good stability.
		}
\end{figure}

\begin{figure}
      \includegraphics[width=3in]{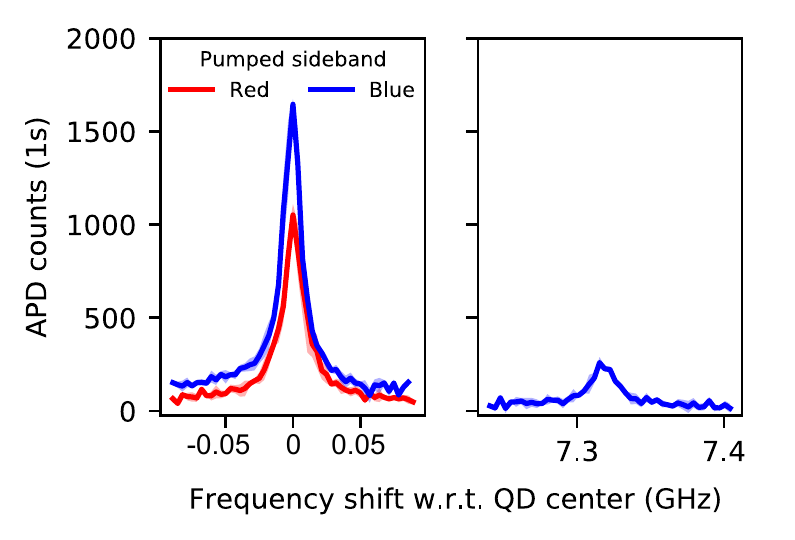}
	\caption{
	Resonance fluorescence sideband measurements under (blue curves) first-order blue-detuned pumping at $+f_0$ and (red curve) red-detuned pumping at $-f_0$ with respect to the QD’s center frequency (set to 0). Microwave driving power was $-26$ dBm at $f_0$=3.65765 GHz. Optical pump power was $\lesssim$300 nW. The large asymmetry in counts between the blue curves around 0 and $2f_0$=7.315 GHz is due to resonant enhancement of the phonon-creating scattering process for blue-detuned pumping in the presence of the QD, verifying operation in the resolved-sideband limit. Counts at $+2f_0$ largely arise from first-order scattering from the Lorentzian tail of the QD and may have a small contribution from other quasi-resonant QDs in the pump spot. The red curve originates from a phonon-annihilating process when pumping the system at $-f_0$.}
\end{figure}

% Surround figure environment with turnpage environment for landscape
% figure
% \begin{turnpage}
% \begin{figure}
% \includegraphics{}%
% \caption{\label{}}
% \end{figure}
% \end{turnpage}

% Create the reference section using BibTeX:
\bibliography{man}

\end{document}